\documentclass{article}
\usepackage{graphicx}

\newcommand{\quer}[1]{\overline{#1}}
\begin{document}
\title{Ground state structure of diluted
antiferromagnets and random field systems}

\author{Alexander K. Hartmann\\
{\small  Institut f\"ur theoretische Physik, Philosophenweg 19, }\\
{\small 69120 Heidelberg, Germany}\\
{\small Tel. +49-6221-549449, Fax. +49-6221-549331}\\
{\small  
hartmann@tphys.uni-heidelberg.de\thanks{permanent address}}\\
{\small and}\\
{\small Theoretische Tieftemperaturphysik, Gerhard Mercator
Universit\"at, }\\
{\small 47048 Duisburg, Germany }}

\maketitle

\begin{abstract}
A method is presented for the calculation of all exact ground states
of diluted antiferromagnets and random field systems in an arbitrary range
of magnetic fields
$B\in [B_{start},B_{end}]$ resp. $\Delta \in [\Delta_{start}, \Delta_{end}]$. 
It works by
calculating  all {\em jump}-fields $B,\Delta$ where the system
changes it's ground state. For each field value  all
degenerated ground states are represented
by a set of (anti-) ferromagnetic clusters and a relation between the 
clusters. So a complete description
of the ground state structure of these systems is possible. 

Systems are investigated up to size $48^3$ on the whole field-range 
and up to $160^3$ 
for some particular fields. The behavior of order 
parameters is investigated, the number of jumps is analyzed 
and the degree of degeneracy  as functions of size and fields is calculated.

{\bf Keywords (PACS-codes)}: random systems (75.10.N), 
ground state calculation (75.40.H), order parameters (75.40.C),
graph theory (02.10). 

\end{abstract}

\section{Introduction}

\begin{sloppypar}
The behavior of diluted Ising antiferromagnets in a homogeneous field
(DAFF) and of random field Ising ferromagnets (RFIM) is still not
well understood (for a review see \cite{reviews}). Although it has
been argued that DAFF and RFIM belong to the same universality classes
\cite{fishman, cardy}
recent results \cite{cambier, nowak1, nowak2, esser} suggest that there
might be essential differences.
\end{sloppypar}

In \cite{ogielski} integer optimization algorithms where 
applied to the RFIM in order to calculate 
exact ground states at fixed fields in
polynomial time.
In \cite{alex1} different 
degenerated ground states of DAFFs and RFIMs were calculated
by using a simple extension of these optimization methods.

The invention of new fast algorithms \cite{traeff} inspired us to develop
an algorithm which calculates the ground states for all fields and
is presented in this paper. Additionally the complete degenerated structure of
a ground state is computable at once, by the use of some more 
complex algorithms
from graph theory \cite{picard2} . Applying these methods a far better
insight into the nature of ground states of disordered systems is possible. 

All systems we investigated were cubic $LxLxL$ lattices 
of Ising spins $\sigma_i=\pm 1$ with 
periodic boundary conditions. For  a DAFF each
lattice site is occupied ($\epsilon_i=1$) with probability $p$ (otherwise
$\epsilon_i=0$). The DAFF is described by the following Hamiltonian

\begin{equation}
H = J\sum_{<i,j>}\epsilon_i \epsilon_j \sigma_i\sigma_j - 
B\sum_i \epsilon_i \sigma_i
\end{equation}

with the uniform external field $B>0$ and the interaction constant $J>0$. The
sum $<i,j>$ runs over all pairs of nearest neighbors.
For low temperatures and small fields $B<B_c$ the DAFF has an
antiferromagnetic phase. In this paper we show for the critical field
$B_c \approx 0.9$. For finite
fields and low temperatures  exists a frozen domain state which
is characterized by fractal domains \cite{nowak1, esser, usadel}. For higher 
fields or higher temperatures the DAFF is paramagnetic.

The Hamiltonian for the RFIM is
\begin{equation}
H = -J\sum_{<i,j>} \sigma_i\sigma_j - \sum_i B_i\sigma_i
\end{equation}

All sites ore occupied, but the magnetic fields $B_i$ are site dependent 
and distributed according a bimodal ($\pm\Delta$-RFIM) or a Gaussian
(Gaussian-RFIM) probability distribution. 
Their probability density functions are

\begin{eqnarray}
p(B_i) & = & \frac{1}{\sqrt{2\pi}\Delta}e^{-(B_i/\sqrt{2}\Delta)^2} \\
p(B_i) & = & \frac{1}{2}(\delta(B_i-\Delta) + \delta(B_i+\Delta))
\end{eqnarray}

Like the DAFF the 3d RFIM has a long range ordered low temperature and
low field phase, a disordered phase for finite fields (critical field:
$\Delta_c =2.35$ \cite{ogielski}) and is paramagnetic for high temperatures
or high fields.

The outline of the paper is as follows. The first section explains how all 
degenerated ground states of a system at a fixed field value
can be represented by a relation of clusters.
The second section gives an algorithm which calculates the ground states
for all fields. In the third section we show the results of our calculations.
We present the order parameter for sample systems of different sizes,
estimate the values for the critical fields,
analyze how the number of jumps increases with increasing system
size and investigate the degree of degeneracy as function of field and size.
In the last section we summarize our results.

\section{Calculation of a ground state}

We used well known algorithms from graph theory \cite{swamy, claibo, knoedel}
to calculate the
ground state of a system at given field $B,\Delta$ and interaction
constant $J$. All quantities $B, \Delta, J$ have to be integer-valued.
The calculation works by transforming the system into a network
\cite{picard1}, and calculating the maximum flow \cite{traeff, tarjan}. 
\footnote{Implementation details: We used Tarjan's wave algorithm together
with the heuristic speed-ups of Tr\"aff. In the construction of 
the {\em level graph} we allowed not only edges $(v, w)$
with level($w$) = level($v$)+1, but also all edges $(v,t)$ where $t$
is the sink. For this measure, we observed an additional speed-up of
roughly factor 2 for the systems we calculated.} 
All degenerated ground states of the system are given \cite{picard2} 
by a set $\quer{V} = \{S=V_0, V_1, \ldots, V_n, V_{n+1}=T\}$ of
clusters, and a binary
relation $\quer{R}$ defined on
$\quer{V}$. Each cluster $V_i$ is a set of
antiferromegnetically (DAFF) resp. ferromagnetically (RFIM) ordered spins. 
These spins are not necessarily spatial connected.

From the pair $(\quer{V}, \quer{R})$ a ground state is calculated in the
following way. Each cluster $V_k$ ($k=0,\ldots,n+1$) is assigned a {\em
orientation} (or {\em sign}) $s(V_k)$: 

The signs of the clusters $S, T$
are determined: $s(S) = 1$, $s(T) = -1$. For all other clusters 
$V_k$ ($k=1,\ldots,n$), called {\em inner clusters}, the
sign can be chosen free ($s(V_k)=\pm 1$) under consideration of the
condition ($l=1,\ldots,n$): 

\begin{equation}
\mbox{if}\,\,(s(V_k) = +1 \,\,\mbox{and}\,\, (V_k,V_l)\in\quer{R})\,\,
\mbox{then}\,\,s(V_l) = +1 
\end{equation}

I.e., if a cluster has the sign $+1$, then all successors in the relation
$\quer{R}$ must have the sign $+1$ too. 

From the signs of the clusters the spin states of the spins $i\in V_k$ 
are calculated by

\begin{eqnarray}
\sigma_i = & t(i) s(V_k) & (\mbox{DAFF}) \\
\sigma_i = & s(V_k) & (\mbox{RFIM}) 
\end{eqnarray}

The function $t(i)$ divides the lattice of the DAFF into two sublattices:
$t(i) = (-1)^{x+y+z}$ for a spin $i$ with coordinates $(x,y,z)$.
So the clusters $S, T$ consist of the spins with fixed states, for
each direction one cluster, all other clusters contain spins which
contribute to the degeneracy of the ground state. The dependencies
between spins of adjacent clusters are given by the relation $\quer{R}$

As example we consider an one dimensional RFIM chain with open ends
consisting of four spins $\sigma_1,\ldots,\sigma_4$. The random fields
are $B_1=2, B_2=B_3=0, B_4=-2$ ($J=1$). So the states of the end-spins
are fixed in the ground state $\sigma_1=1, \sigma_4=-1$. The states of the 
inner spins are not fixed. The system is visualized in figure 
\ref{example_sys}.

The energy of the system as function of the two inner spins is 
$H=-4 -\sigma_2 + \sigma_3 - \sigma_1\sigma_2$. The energies for the
four possible states are shown in the following table.

\begin{center}
\begin{tabular}{rr|c}
$\sigma_2$ & $\sigma_3$ &  H \\ \hline
$+1$ & $+1$ & -5 \\
$+1$ & $-1$ & -5 \\
$-1$ & $+1$ & -1 \\
$-1$ & $-1$ & -5
\end{tabular}
\label{tab_four_states}
\end{center}

So the ground state of the system is threefold degenerated. It can be 
described
by the rule: The inner spins are not fixed, but if $\sigma_3=+1$ then
$\sigma_2=+1$ must hold too. So the ground state is formally given by

\begin{equation}
\quer{V}=\{S,V_1,V_2,T\}, S=\{1\}, V_1=\{2\}, V_2=\{3\}, T=\{4\}
\end{equation}

and

\begin{equation}
\quer{R} = \{(2,1)\}
\end{equation}

\begin{figure}[ht]
\begin{center}
\scalebox{0.65}[0.75]{\rotatebox{-90}{\includegraphics{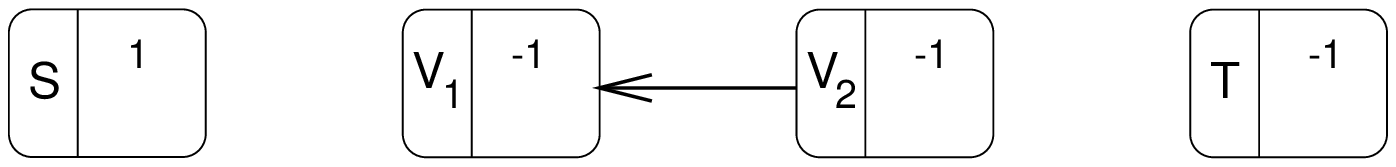}}}
\end{center}
\caption{}
\label{example_rr}
\end{figure}

The ground state can be displayed as a graph. The nodes represent the
clusters and the edges represent the relation. Because for 'real' systems
the clusters are too large, only the order parameter 
of each cluster is given in the
graph, instead of enumerating all spins. For the inner clusters, where the
order parameter can take two values, the value for the cluster orientation
$s(V_k)=-1$ is displayed. 
So the example system is represented by the graph shown in figure
\ref{example_rr}.

\section{Calculation of ground states for all $B,\Delta$}

Let's examine a Cluster $C$ in a DAFF with magnetization 
$m_C=m_C^{\uparrow}-m_C^{\downarrow}$ ($m_C^{\uparrow},m_C^{\downarrow}\ge 0$
denote the number of spins in up and down direction). 
The cluster is supposed to exist as a coherent unit 
for an finite $B$-interval $[B_1,B_2]$. Without loss of generality
all other clusters can be
supposed to keep their orientations inside this interval.

The cluster is
connected to its environment by $N_C^s$ satisfied and $N_C^u$ unsatisfied 
bonds. The
part of the cluster energy describing its connection to its surface
and to the external
field  is

\begin{equation}
E_c(B) = (N_C^u-N_C^s)J - m_CB = (N_C^u-N_C^s)J- 
(m_C^{\uparrow}-m_C^{\downarrow})B
\label{clusterenergy1}
\end{equation}

($N_C^u,N_C^s,m_C^{\uparrow},m_C^{\downarrow}$ depend itself on $B$.)
For a ground state $E_c(B)\rightarrow \min$ holds.
Let's suppose that the cluster flips its orientation at 
$B=B^j \in [B_1,B_2]$. That implies that the magnetization of the
cluster is nonzero, because the environment of the cluster is supposed
to be unchanged, and otherwise there would not occur a jump. Since
the number of satisfied/unsatisfied bonds and the number of up/down
spins exchange, when a cluster is flipped, the energy of a flipped cluster
changes its sign. That implies $E_c(B^j)=0$, because $E_c(B) \le 0$ for
all $B\in [B_1,B_2]$. So the
cluster can take both orientations  at $B=B^j$. That leads to

\begin{equation}
B^j/J = \frac{N_C^u-N_C^s}{m_C^{\uparrow}-m_C^{\downarrow}}
\end{equation}

($N_C^u,N_C^s,m_C^{\uparrow},m_C^{\downarrow}$ all from $B<B^j$ or all
from $B>B^j$.)
Because the cluster $C$ is part of cluster $S$ or $T$, these values are
not directly accessible,
we want to express this ``jump-field'' $B^j$ using characteristic
values of the graphs $\overline{R}$ immediately ``before'' ($B<B^j$) and
``after'' ($B>B^j$) the jump. Because the cluster does not contribute
to the degeneracy (which holds only for clusters with zero
magnetization and $N_C^u=N_C^s$) it moves during the jump from the
(super)cluster $S$ to $T$ or vice versa. So we get ($m_S, m_T$:
magnetizations of $S,T$)

\begin{equation}
|m_C^{\uparrow}-m_C^{\downarrow}| = |m_T(B<B^j)-m_T(B>B^j)|\;\; 
\left(= |m_S(B<B^j)-m_S(B>B^j)|\right)
\end{equation}

Using $N_C^u(B<B^j)-N_C^s(B<B^j) = N_C^u(B<B^j) - N_C^u(B>B^j)$ 
$= N^u(B<B^j)-N^u(B>B^j)$ (the rest of the system does not change,
$N^u$ = total number of unsatisfied bonds in
the system) and because $B^j/J>0$ we have as result:

\begin{equation}
B^j/J = \frac{|N^u(B<B^j)-N^u(B>B^j)|}{|m_T(B<B^j)-m_T(B>B^j)|}
\label{jumpfield1}
\end{equation}


For a RFIM the formula describing the relevant terms of the energy of
a cluster (similar to (\ref{clusterenergy1})) is

\begin{equation}
E_c(\{B_i\}) = (N_C^u-N_C^s)J- \sum_{i\in C} B_i \sigma_i
\end{equation}

A given realization of a RFIM is determined by the local field values
$b_i:= B_i(\Delta=1)$, so $B_i=\Delta b_i$. For the $\pm\Delta$-RFIM
the $b_i$ is just the {\em sign} of the random field. Defining
the total random field sign of the cluster by
$\Sigma_C := \sum_{i\in C}b_i$ we get

\begin{equation}
E_c(\Delta)= (N_C^u-N_C^s)J - \Sigma_C s(C) \Delta
\end{equation}

Comparing with (\ref{clusterenergy1}) we see that $\Sigma_C s(C)$ plays
the role of the magnetization and $\Delta$ is equivalent to the
external field $B$ of a DAFF. A RFIM cluster moves at a jump from the
super-cluster $S$ to $T$ or vice versa, so we have 
$|\Sigma_C| = |\Sigma_T(\Delta<\Delta^j)-\Sigma_T(\Delta>\Delta^j)|$ and 
$|s(C)|=1$. 
So the field $\Delta^j$
where the cluster flips is calculated by:

\begin{equation}
\Delta^j/J = \frac{|N^u(\Delta<\Delta^j)-N^u(\Delta>\Delta^j)|}
                  {|\Sigma_T(\Delta<\Delta^j)-\Sigma_T(\Delta>\Delta^j)|}
\label{jumpfield2}
\end{equation}

\begin{figure}[ht]
\begin{center}
\scalebox{0.65}[0.75]{\rotatebox{-90}{\includegraphics{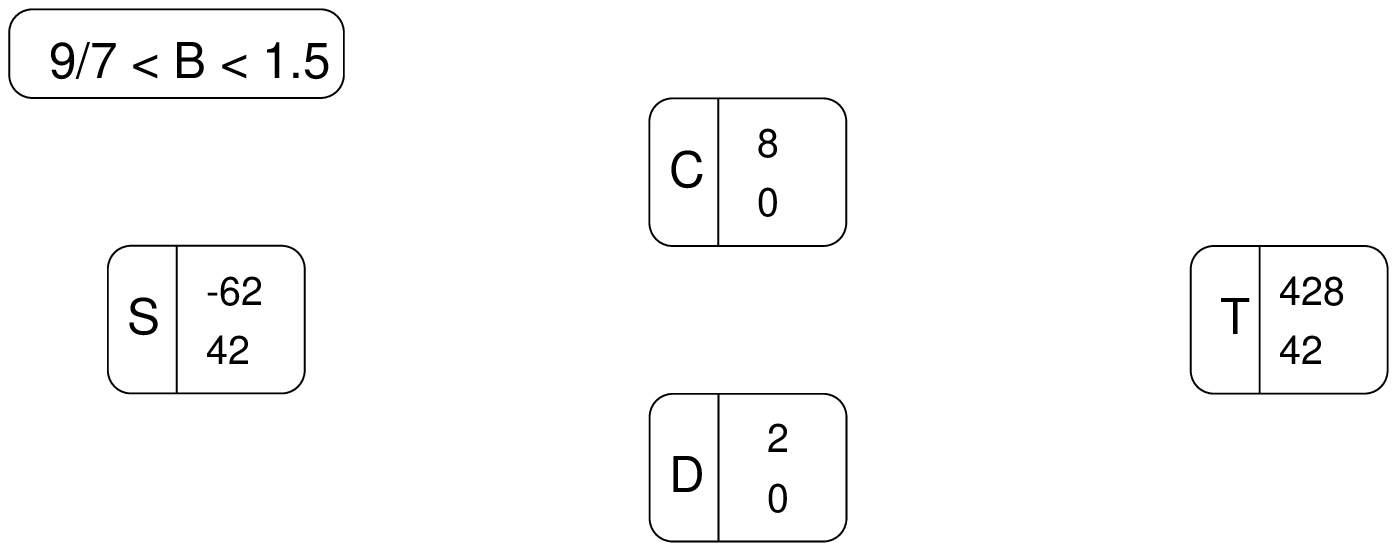}}}
\end{center}
\caption{}
\label{rr_jump1}
\end{figure}

\begin{figure}[ht]
\begin{center}
\scalebox{0.65}[0.75]{\rotatebox{-90}{\includegraphics{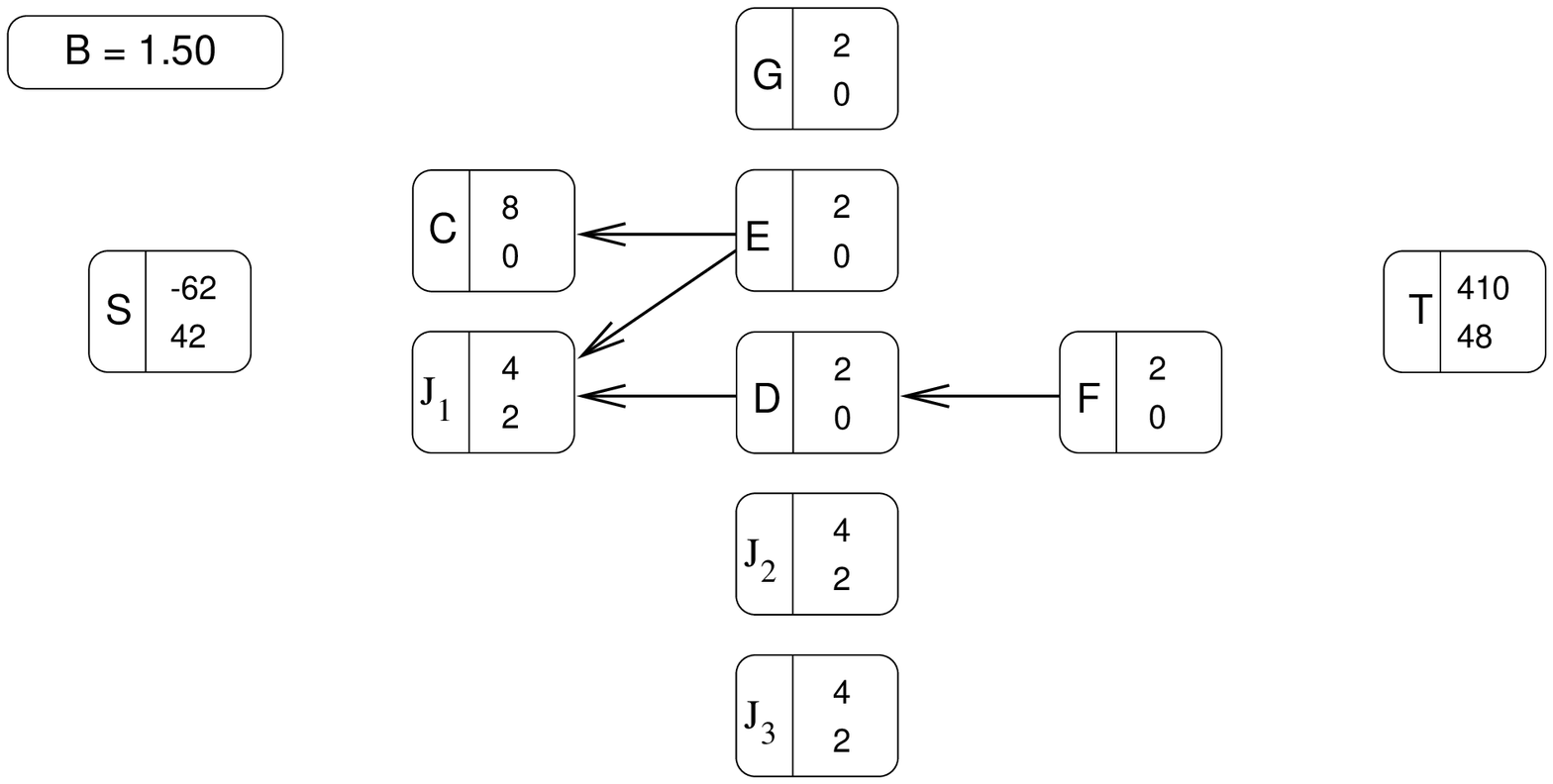}}}
\end{center}
\caption{}
\label{rr_jump2}
\end{figure}

\begin{figure}[ht]
\begin{center}
\scalebox{0.65}[0.75]{\rotatebox{-90}{\includegraphics{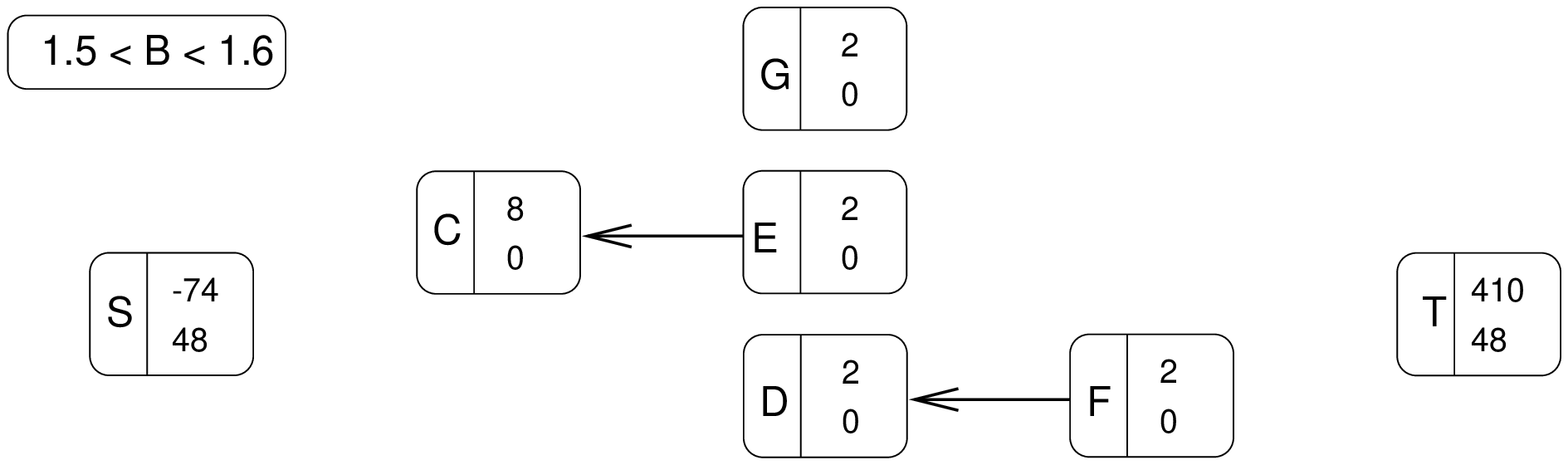}}}
\end{center}
\caption{}
\label{rr_jump3}
\end{figure}

For illustration
in Figures \ref{rr_jump1} to \ref{rr_jump3} the ground states of a
sample DAFF system ($L=10$, $p=0.5$) are displayed 
for $9/7<B<1.6$. In addition to the
order parameter of each cluster (upper value) its magnetization
(lower value) is given.

A cluster with zero magnetization and equal number of satisfied and
unsatisfied bonds can take both orientations over
a range of external fields.
For $9/7<B<1.5$ these holds for clusters $C, D$.
 The degeneration of the system has
degree $2x2=4$, because each cluster $C, D$ can take independently two
orientations. 

At $B^j=1.5$ a cluster $J$ (an aggregation of the small clusters
$J_1,J_2,J_3$) reverses its orientation. Here the number of total
unsatisfied bonds is $N^u(9/7<B<1.5)=39$ and $N^u(1.5<B<1.6)=48$ (not
visible from the graphic representation, but easily calculated using
the spins from clusters $S$, $T$, because all other clusters have
$N_C^u=N_C^s$). The
magnetization of $T$ changes from $m_T(B<B^j)=42$ to $m_T(B>B^j)=48$
(figure \ref{rr_jump3}).
Using formula (\ref{jumpfield1}) we get 
$B^j/J=\frac{|48-39|}{|48-42|}=\frac{9}{6}=1.5$

\begin{sloppypar}
Each of the clusters $G,J_1,J_2$ can take independently two
orientations (={\em states}). The clusters $C, E$ alone can take 
three combinations of orientations  because
$s(E)=+1$ implies $s(C)=+1$. The clusters $J_1, D, F$ alone can take 4
states. Because $s(E)=+1$ implies $s(J_1)=+1$ we have for all 5
clusters together $4+4+(4-1)=11$ states. So the system is $2^3x11=88$
fold degenerated.

For $1.5<B<1.6$ the cluster $J$ has been absorbed by the cluster $S$.
Now the system has a degree of degeneracy of $2x3x3=18$. Note that the
environment of the cluster $J$ has changed during the jump as well. But
only clusters $V$ with $m_V=0$ and $N_V^u=N_V^s$ are affected. So
the formula for $B^j/J$ is still correct.
\end{sloppypar}

Using the formulas (\ref{jumpfield1}, \ref{jumpfield2}) the complete
behavior
of a DAFF/RFIM can be easily calculated. We show how the procedure
works for a DAFF, the RFIM is similar. One starts with a certain
value for $B=B_0$ (There must no jump occur, otherwise the formulas are
not valid) and calculates the
graph $\overline{R}(B_0)$. Then $B$ is increased in macroscopic
steps $\Delta B$ 
up to the value $B_1$ where the calculated graph $\overline{R}(B)$
differs from $\overline{R}(B_0)$ the first time. Then the jump-field
$B_i$ can be calculated from the graphs at $B_0,B_1$. One has to check
if the jump is really the first jump occurring for $B>B_0$. This is
done by calculating the graph $\overline{R}(B^j_-)$ for a field
$B^j_-$ which is infinitesimal smaller than $B^j$ and comparing with
$\overline{R}(B_0)$. If they are the same, the jump $B^j$ is really the
first for $B>B^j$. If not, a new jump-field $B^{j'}$ is calculated
using $\overline{R}(B_0)$ and $\overline{R}(B^j_-)$ and at $B^j$ may not be a
jump at all. This procedure can
be iterated until the jump next to $B_0$ is found. 

It can be easily seen that this algorithm really produces the next
jump starting from $B_0$. Suppose we have two jumps $B^j_1 <B^j_2$
between $B_0$ and $B_1$. Then $B^j$ is given by
(for a DAFF)

\begin{equation}
B^j/J = \frac{|N^u(B_0) - N^u(B_1)|}{|m_T(B_0)-m_T(B_1)|}
\end{equation}

Because $N^u$ and $m_T$ are monotonic in $B$, the effect on that values
of the both jumps sum up. So we have ($B^j_+$ slightly larger than
$B^j$)

\begin{equation}
B^j/J = \frac{|N^u(B_0) - N^u(B^j_-)| + |N^u(B^j_+) - N^u(B_1)|}
           {|m_T(B_0) - m_T(B^j_-)| + |m_T(B^j_+) - m_T(B_1)|}
\end{equation}

From $B^j_1<B^j_2$ and because $n_1/d_1 < n_2/d_2$ implies $n_1/d_1 <
(n_1+n_2)/(d_1+d_2) < n_2/d_2$ we get

\begin{equation}
B^j_1/J = \frac{|N^u(B_0)-N^u(B^j_-)|}{|m_T(B_0)-m_T(B^j_-)|} < B^j/J
< \frac{|N^u(B^j_+)-N^u(B_1)|}{|m_T(B^j_+)-m_T(B_1)|} = B^j_2/J
\end{equation}

By induction follows that the algorithm finds the next jump from $B_0$
 for all number of jumps
between $B_0$ and $B_1$. The next jump from a given jump $B^j$ is found
by restarting the algorithm at $B_0 := B^j_+$. Beside this linear search
a bisection method starting with $B_0=B_{start}$ and $B_1=B_{end}$
should be possible. But we implemented the linear search, because it is
more straight forward. Taking into account,
that the calculation of the graphs $\overline{R}$ requires $B,J$ to be
integers, the linear search of all states $\overline{R}$ in a range
$[B_{start}/J_0,B_{end}/J_0]$
can be algorithmically displayed as follows:

\newpage

\newlength{\tablen}
\settowidth{\tablen}{xxx}
\newcommand{\tabspace}{\hspace*{\tablen}}
\begin{tabbing}
\tabspace \= \tabspace \= \tabspace \= \tabspace \= \tabspace \=
\tabspace \= \kill
{\bf algorithm} all ground states($B_{start},B_{end},J_0$)\\
{\bf begin}\\
\> $B \leftarrow B_{start}$\\
\> $\Delta B \leftarrow J_0$\\
\> stop $\leftarrow$ {\bf false}\\
\> {\bf while} ( {\bf not} stop )\\
\> {\bf begin}\\
\>\> $B_0 \leftarrow B$\\
\>\> calculate $\overline{R}(B_0)$\\
\>\> same\_R $\leftarrow$ {\bf true}\\
\>\> {\bf while}(same\_R)  /* Search for first state different from
                         $\overline{R}(B_0)$ */\\
\>\> {\bf begin}\\
\>\>\> {\bf if}(same\_R {\bf AND} $B>B_{end}$) {\bf then} \\
\>\>\>\> stop $\leftarrow$ true; \\
\>\>\> {\bf else} \\
\>\>\> {\bf begin} \\
\>\>\>\> $B \leftarrow B+\Delta B$ \\
\>\>\>\> calculate $\overline{R}(B)$ \\
\>\>\>\> $\Delta B \leftarrow 2*\Delta B$\\
\>\>\>\> {\bf if}( $\overline{R}(B_0) \neq \overline{R}(B)$) {\bf then} \\
\>\>\>\>\> same\_R $\leftarrow$ false \\
\>\>\> {\bf end}\\
\>\> {\bf end} \\
\>\> {\bf if}( stop) {\bf then} \\
\>\>\> {\bf continue}\\
\>\> found $\leftarrow$ false\\
\>\> $B_1 \leftarrow B$\\
\>\> {\bf while}( {\bf not} found) /* search for next jump */ \\ 
\>\> {\bf begin}\\
\>\>\> $B^j_- \leftarrow (\mbox{int}) \left[
\frac{| N^u(B_0) - N^u(B_1)|}{|m_T(B_0)-m_T(B_1)|}*J_0 +0.5 \right] - 1$\\
\>\>\> calculate $\overline{R}(B^j_-)$\\
\>\>\> {\bf if}( $\overline{R}(B^j_-) = \overline{R}(B_0)$) {\bf begin}\\
\>\>\>\> found $\leftarrow$ {\bf true}\\
\>\>\> {\bf else}\\
\>\>\>\> $B_1 \leftarrow B^j_-$ \\
\>\> {\bf end}\\
\>\> There is a jump at $B^j \leftarrow | N^u(B_0)-N^u(B_1)|,\;
      J^j\leftarrow|m_T(B_0)-m_T(B_1)|$\\
\>\> $B \leftarrow (\mbox{int}) \left[ J_0*B^j/J_j+0.5\right] +1$\\
\>\> $\Delta B \leftarrow$ difference between last two jumps \\
\> {\bf end}\\
{\bf end}\\
\end{tabbing}

Technical remarks: 
\begin{itemize}
\item $J_0$ should be chosen sufficiently large. For the
DAFF and the $\pm\Delta$-RFIM $J_0>N$ guarantees to find all
jumps. Since the values for $b_i$ can take all real values for the
Gaussian RFIM there is always a small probability of missing 
jumps. So we rounded the $b_i$-values to three significant digits and
used $J_0 > 10^3N$.
\item The field increment $\Delta B$ used for searching the next state
which is different from $\overline{R}(B_0)$ is set to the difference between
the last two jumps found (initially $J_0$). During the search $\Delta B$
is doubled in each iteration, so jumps lying close to each other do not
slow down the following calculations.
\item The states $\overline{R}(B_0), \overline{R}(B_1),
\overline{R}(B^j_-)$ must not be exactly at jumps in order to guarantees
that the algorithm works. This is confirmed by checking if there are
 inner clusters with nonzero magnetization. If this is true the
field $B$ is in-/decrease about 1 up to a non-jump field (not shown in the 
algorithmic representation). For the
DAFF/ $\pm\Delta$-RFIM this is guaranteed to be done in one step.
\item For the DAFF/$\pm\Delta$-RFIM the states $\overline{R}$ at the
jumps $(B^j,J_j)$ can be calculated. For the Gaussian-RFIM that is not always
possible directly, because the values of $B_i$ are rounded off.
\end{itemize}

\section{Results}

We performed calculations of ground states over  large 
$B,\Delta$-intervals (values are related to $J=1$) for 
system sizes $L=8$ to $L=48$ for the DAFF ($p=0.5$), $\pm \Delta$-RFIM and the 
Gaussian-RFIM. The number of samples per system type and size are shown
the following table.

\begin{center}
\begin{tabular}{|c||c|c|c|}
\hline
L & \begin{minipage}{3cm} \center DAFF \\ $B \in [0,6]$ \end{minipage} & 
\begin{minipage}{3cm} 
\center $\pm\Delta$-RFIM\\ $\Delta\in [0,6]$ \end{minipage} & 
\begin{minipage}{3cm} 
\center Gaussian-RFIM \\ $\Delta\in [0,4]$ \end{minipage} 
\\ \hline
8 & 850 & 450 & 250 \\
12 & 850 & 450 & 120 \\
16 & 450 & 250 & 58  \\
24 & 450 & 177 & 32 \\
32 & 50 & 43 & 16 \\
48 & 50 & - & - \\ \hline
\end{tabular}
\label{tab_num_samples}
\end{center}

For the $L=48$ samples we reduced the $B$-interval to $[0.5,1.5]$. For the
$L=32$ Gaussian-RFIM we reduced the $\Delta$-interval to $[1.0,2.8]$. 
 
\begin{figure}[ht]
\begin{center}
\scalebox{0.5}[0.5]{\rotatebox{-90}{\includegraphics{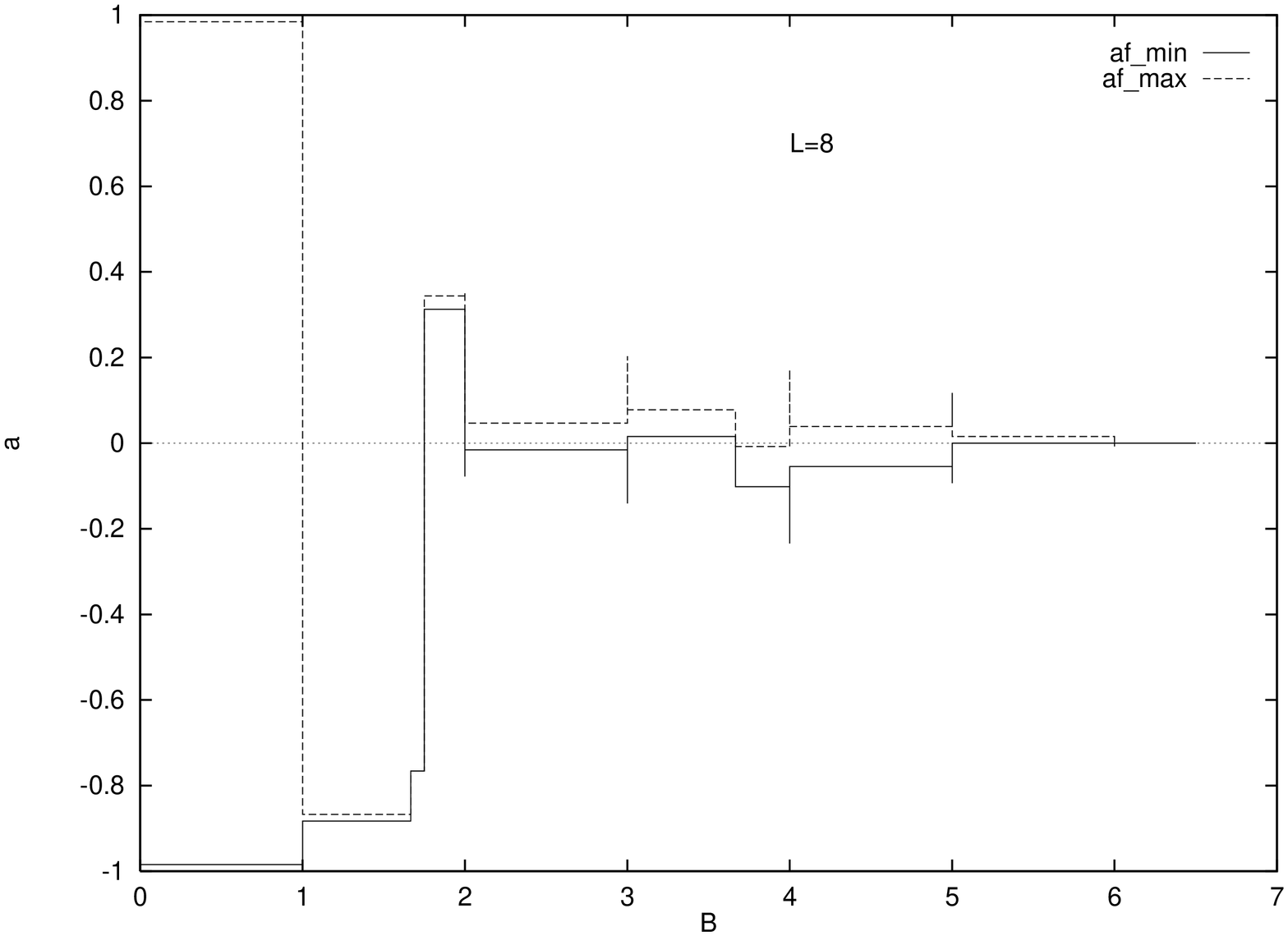}}}
\scalebox{0.5}[0.5]{\rotatebox{-90}{\includegraphics{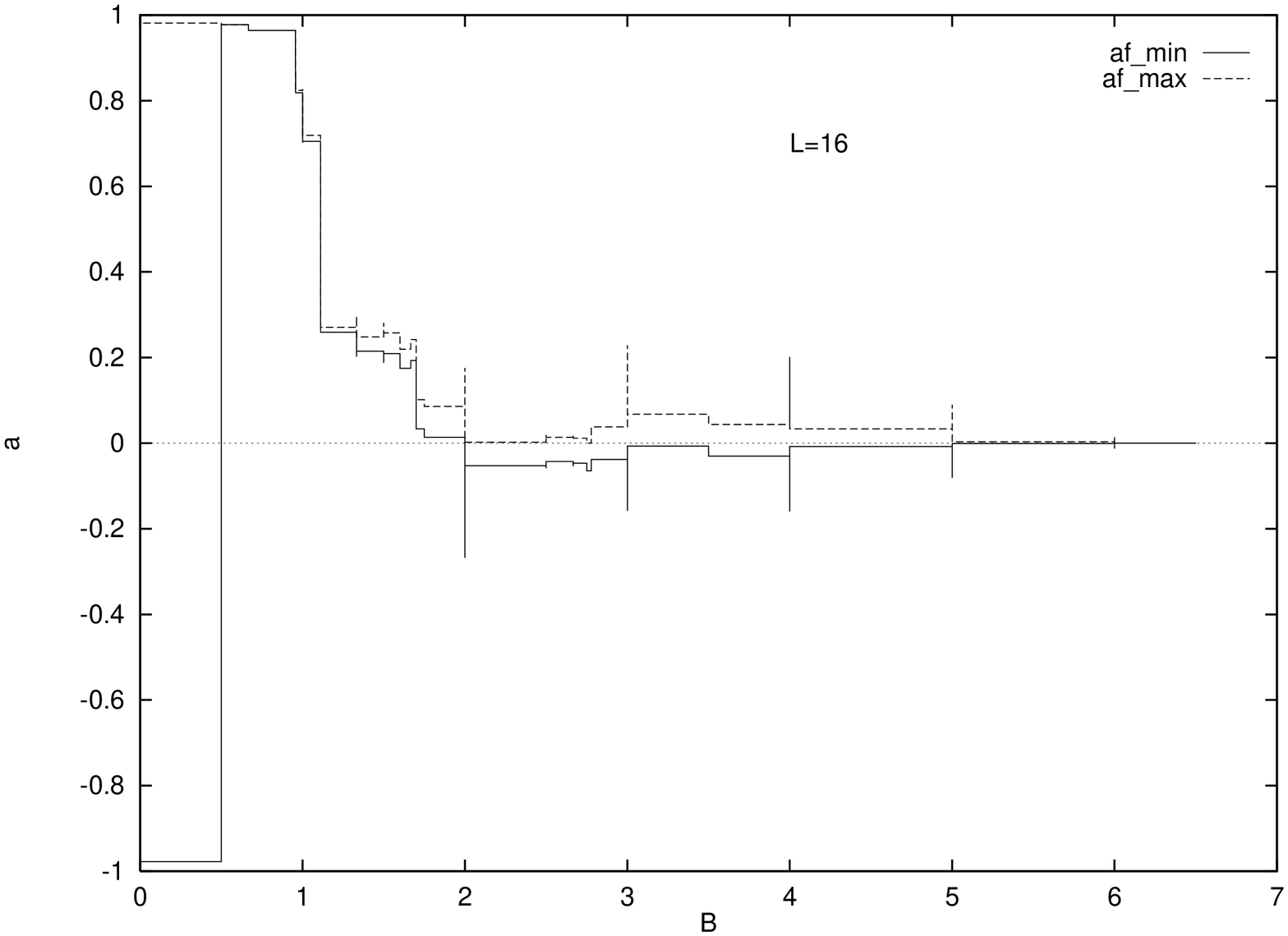}}}
\end{center}
\caption{}
\label{p_af08}
\end{figure}

In figure \ref{p_af08} the antiferromagnetic order parameter per spin

\begin{equation}
a = \frac{1}{N}\sum_{i}\epsilon_i t(i)\sigma_i
\end{equation}

of two DAFF systems with $L=8$ and $L=16$ over the whole range $B \in[0,6.5]$
is displayed. Because of the ground state degeneracy many states are possible.
Here the two extreme values, the order parameter can take, are shown. For the
small system for $B<1$ two states are possible, because its total magnetization
is zero. The larger system has this feature only for $B<0.5$. For both systems
the range of possible $a$-values is relatively small for $1<B<2$ and little
bit larger for $2<B<4$. The range is the largest for integer values $I$ of $B$,
because spins having neighbors with total magnetization $I$ can take both
orientations. 
The main differences between the two systems are: The larger system
exhibits more jumps and the order parameter decreases more quicker than
for $L=8$.

The discrete structure is clearly visible. In both systems
the order parameter is high for small fields and goes stepwise to zero.

Because of the step-structure it is difficult to obtain 'critical'
magnetic fields by calculation of susceptibilities and then use finite
size scaling to go to the $L\to\infty$ limit. Instead we proceed as follows, to
get first informations about how the order parameter changes. We calculated
for each sample the {\em barrier field value} $B_{a <a_0}$ where the
order parameter $a$ falls the first time below $a_0$, i.e.

\begin{equation}
B_{a <a_0} = \min \{B | a(B)<a_0 \}
\end{equation}

\begin{figure}[ht]
\begin{center}
\scalebox{0.5}[0.5]{\rotatebox{-90}{\includegraphics{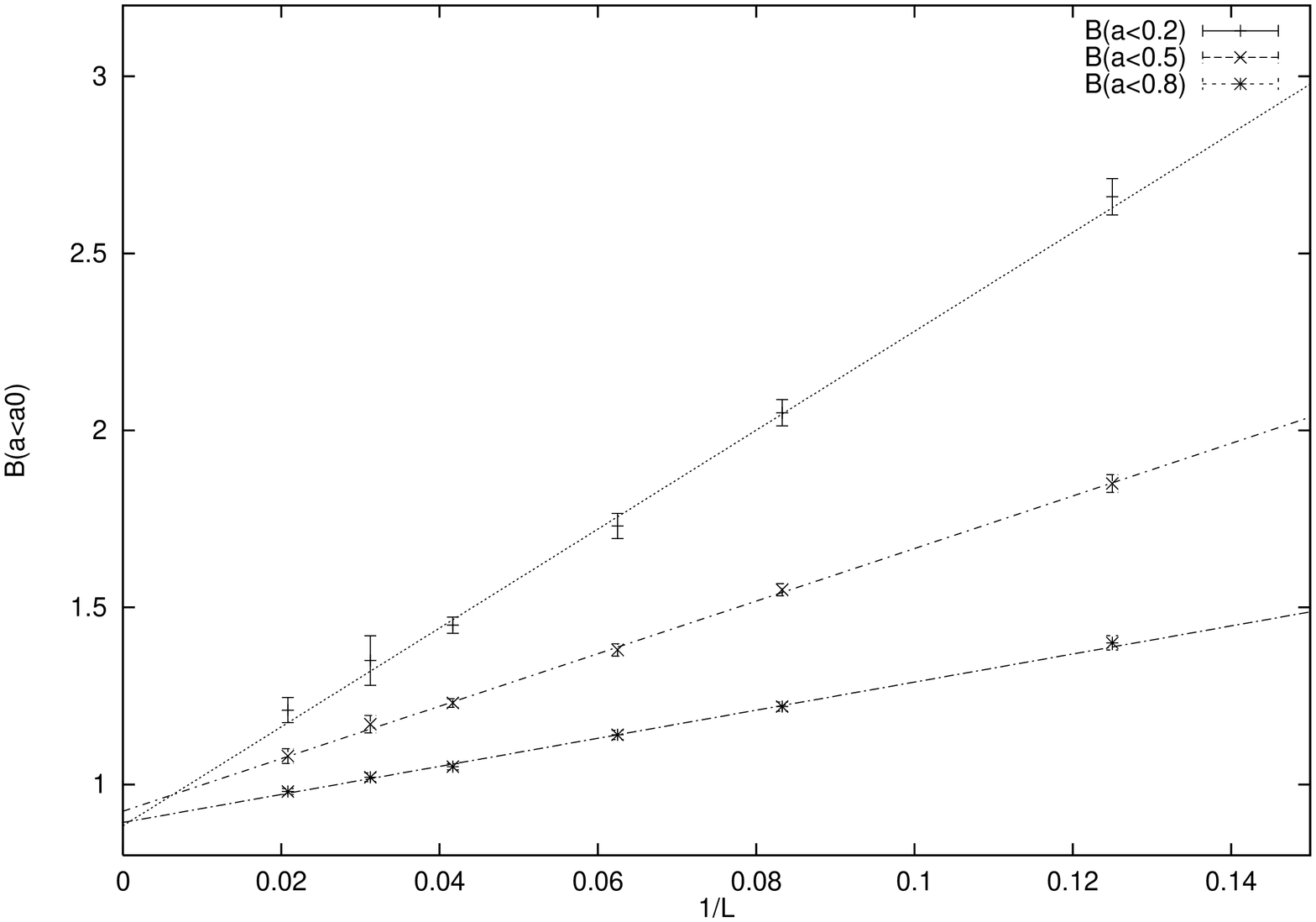}}}
\end{center}
\caption{}
\label{p_af_unter}
\end{figure}

In figure \ref{p_af_unter} the average $B_{a < a_0}$ is plotted against the
inverse system size $1/L$ for $a_0 = 0.2, 0.5, 0.8$. Just to get 
a first impression
we extrapolated to
$L\to\infty$ by fitting straight lines $B_{a<a_0}(1/L) = c/L +B_{a<a_0}(0)$.
 The fit seems to be good, but may have  no physical meaning.
From the figure we can see that $B_{a<0.5}(0) = 0.90 \pm 0.02$, so the 
transition to the domain state occurs for $B_c<1$ !
This is in contradiction to previous results from MC calculations
of systems of size 60x60x61 \cite{nowak1} where $B_c\approx 1.4$ was found.
Because we want to rule out the influence of the fit function,
we checked the result by calculating the ground states for ten $L=160$ 
systems, 
at $B=0.9999$
 where we got an average order parameters of $0.22 \pm0.11$. So the
infinite system has no long range order for $B> 0.9$ !
From figure
\ref{p_af_unter} we can't see, whether the $[B_{a<0.8},B_{a<0.2}]$ interval
remains finite or becomes infinitely small for the $L=\infty$ system. More
calculations and larger systems are needed. 

\begin{figure}[ht]
\begin{center}
\scalebox{0.5}[0.5]{\rotatebox{-90}{\includegraphics{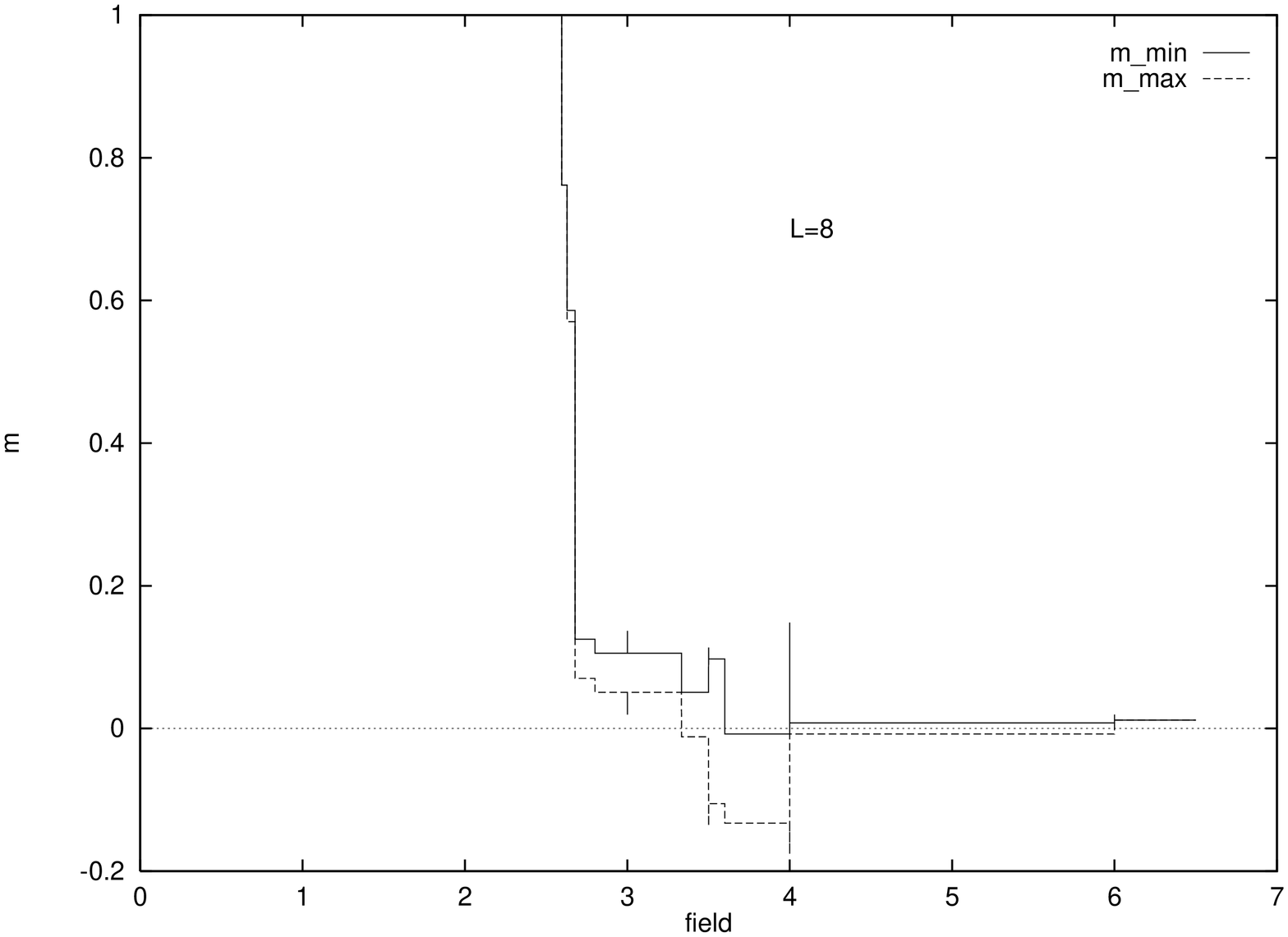}}}
\scalebox{0.5}[0.5]{\rotatebox{-90}{\includegraphics{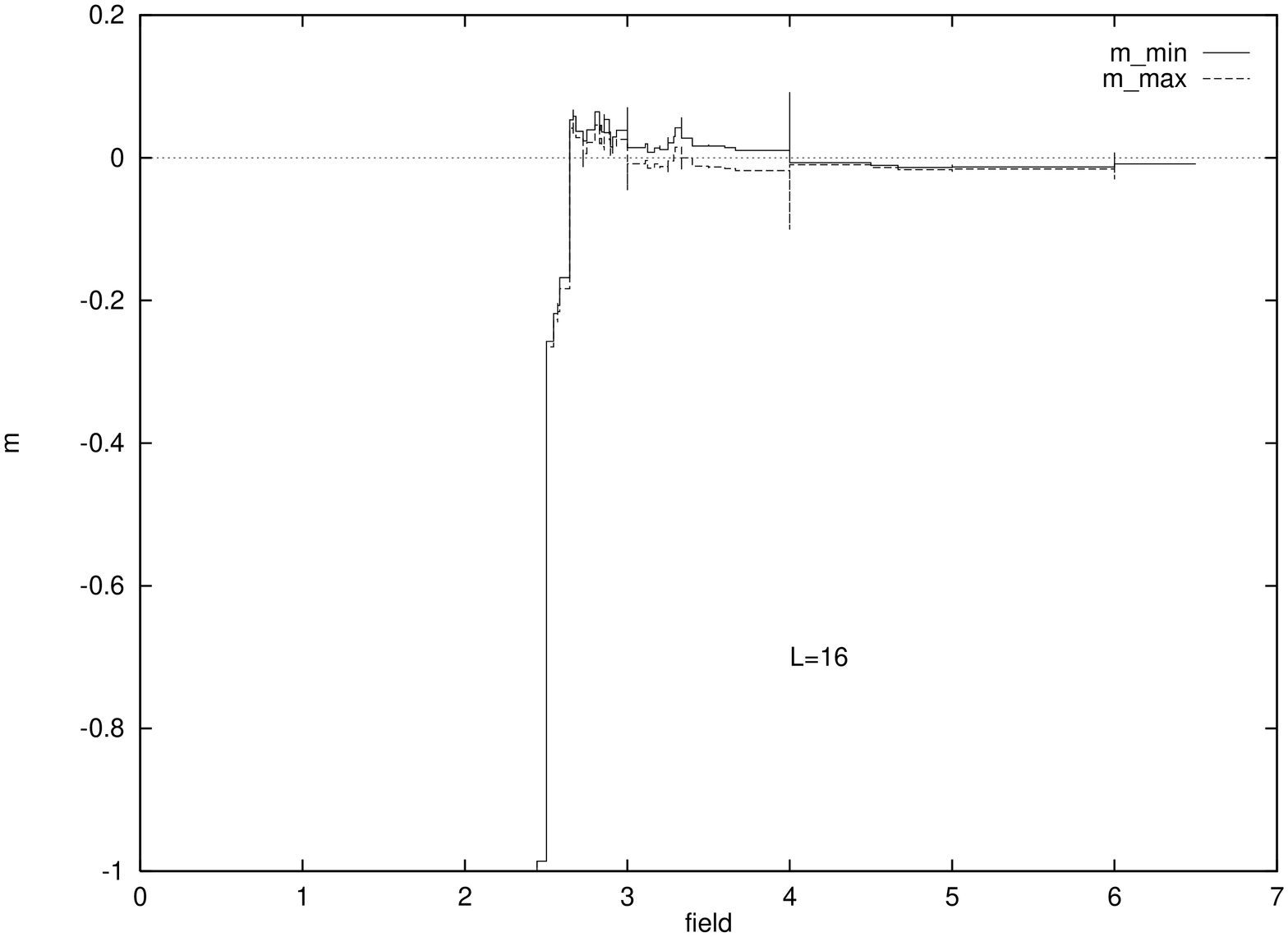}}}
\end{center}
\caption{}
\label{p_rf08}
\end{figure}

In figure
\ref{p_rf08} diagrams are shown for the magnetization per spin

\begin{equation}
m = \frac{1}{N}\sum_i \sigma_i
\end{equation}

of two sample $\pm \Delta$-RFIM.

The impression is similar to the DAFF, but the order parameter goes to
zero at higher fields and there are more jumps in the system because there
are more spins in a RFIM than in a DAFF of the same size $L$. 
Diagrams of Gaussian-RFIM look similar but there are even more jumps
in a system of equal size and the decrease of the order parameter
is less rapid.

\begin{figure}[ht]
\begin{center}
\scalebox{0.5}[0.5]{\rotatebox{-90}{\includegraphics{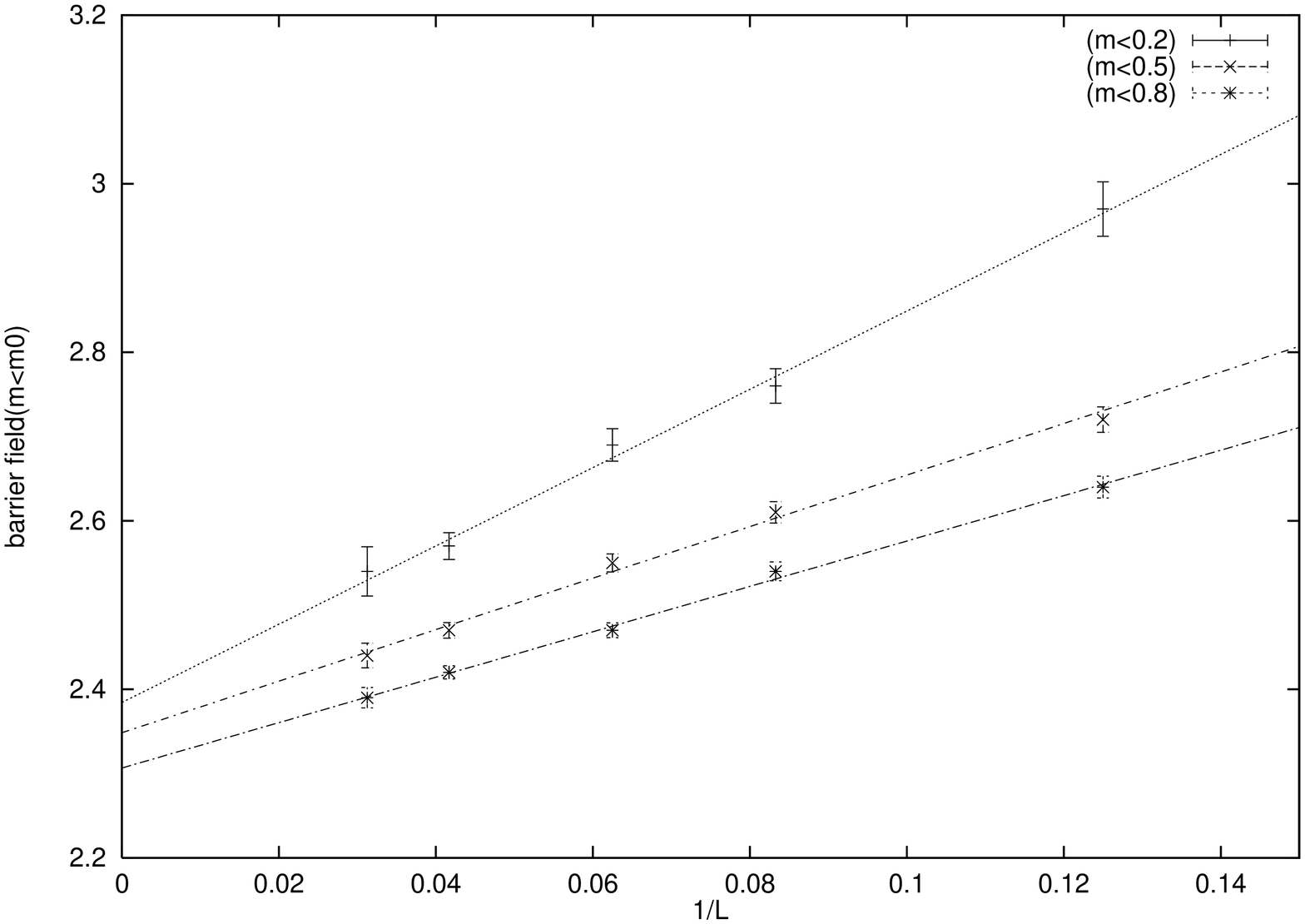}}}
\scalebox{0.5}[0.5]{\rotatebox{-90}{\includegraphics{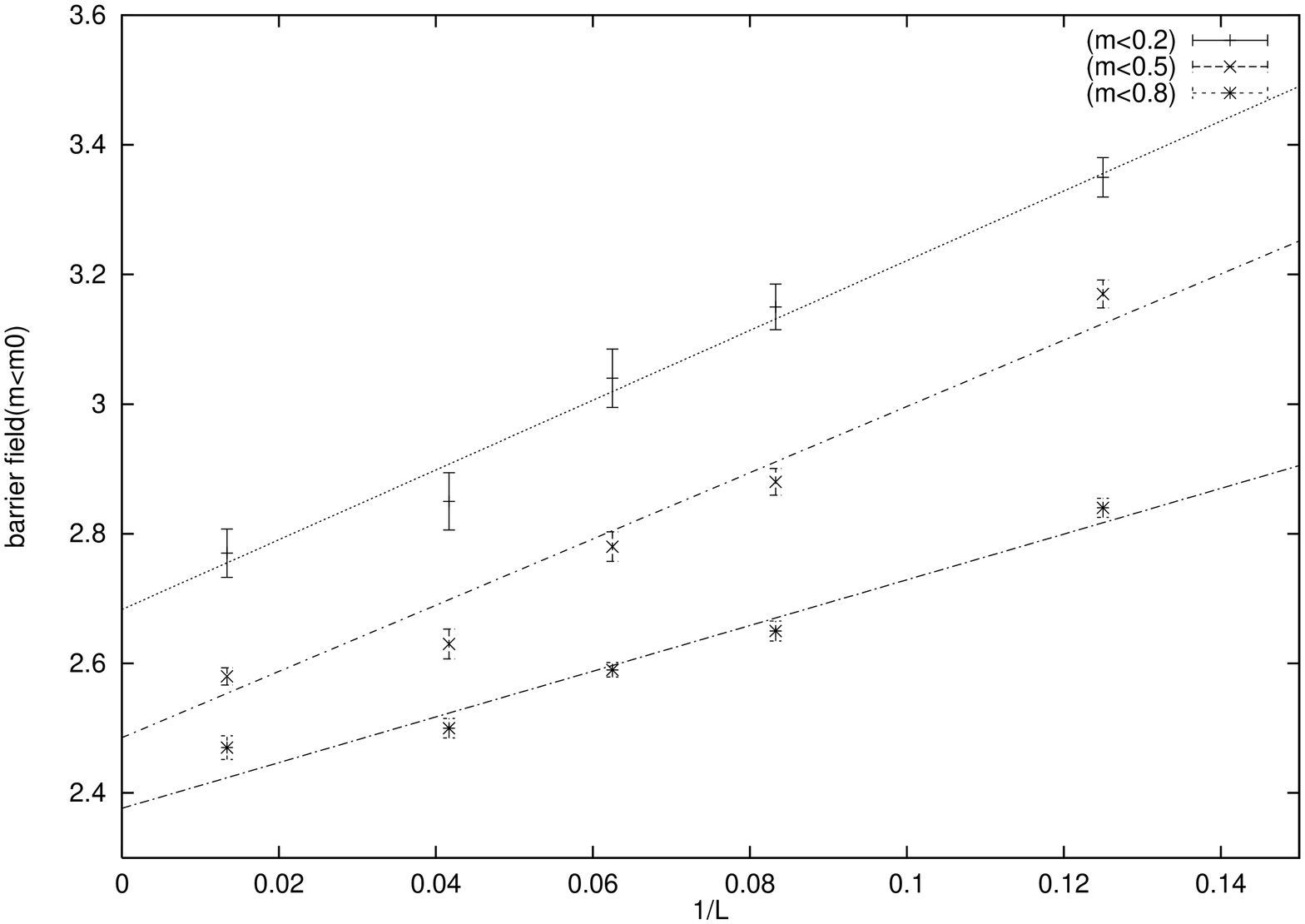}}}
\end{center}
\caption{}
\label{p_rf_unter}
\end{figure}

In analogy to the DAFF we calculated the barrier fields $\Delta_{m<m_0}$ for 
the random field systems. They are display in figure \ref{p_rf_unter}

We got the values $\Delta_{m<0.5}(0) = 2.35\pm 0.01$ ($\pm\Delta$) and
$\Delta_{m<0.5}(0) = 2.49\pm 0.04$ (Gaussian) for the infinite systems.
The value for the $\pm\Delta$ system is consistent with previous
calculations \cite{ogielski}.

For the Gaussian-RFIM the values are less reliable, because we could
calculate only few systems. The Gaussian systems exhibit many jumps and
the number of ground states which have to be calculated is about six times
the number of jumps.
So many ground states have to be calculate to analyze the system, resulting
in more computer time needed.

\begin{figure}[ht]
\begin{center}

\scalebox{0.5}[0.5]{\rotatebox{-90}
         {\includegraphics{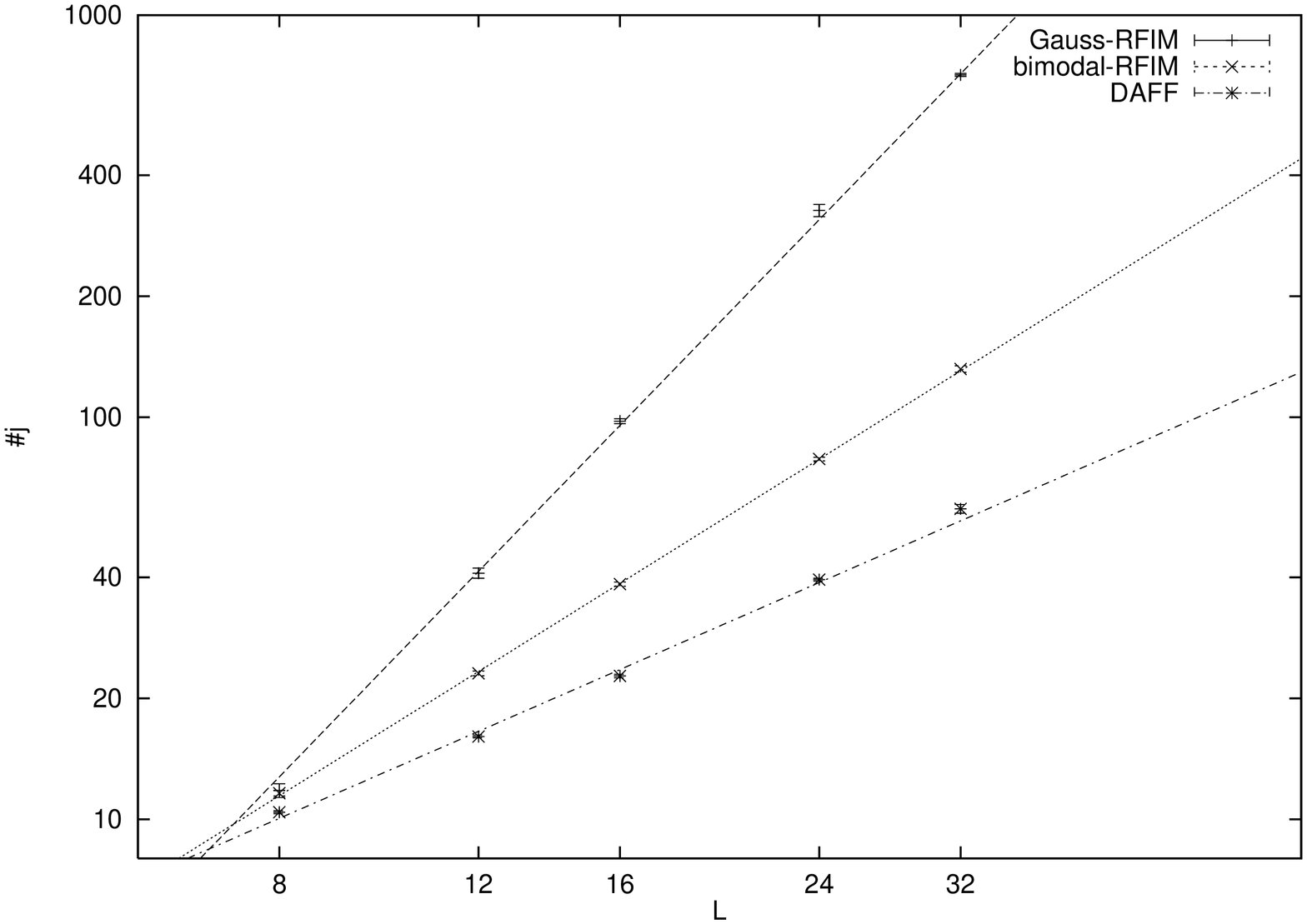}}}
\end{center}
\caption{}
\label{p_numjumps_L}
\end{figure}

For all types of systems the number of jumps $\#j$ increases with system
size. Figure \ref{p_numjumps_L} shows this dependence. The number of jumps
increases roughly like $\#j(L)=c*L^b$ with $b=1.23\pm0.04$ (DAFF), 
$b=1.76\pm0.01$ 
($\pm\Delta$-RFIM), $b=2.91\pm0.03$ (Gaussian-RFIM, only jumps in 
$\Delta\in [1.0,2.8]$ are counted, the number of jumps in [0.0,6.0] is
about four times higher !) This means that the behavior of the system gets 
smoother with increasing size, i.e. the infinite system looses the
discrete structure on finite $B,\Delta$-scales.

We measured also the number of jumps in the $B,\Delta$-interval 
where the order parameter falls from 0.8 to 0.2 for each system.
Even this number increases although the interval
is reduced with increasing system size. We performed a similar
fit to the case above and obtained $b=0.90\pm 0.02$ (DAFF),
$b=1.40\pm 0.06$ ($\pm\Delta$-RFIM) and $b=2.18\pm 0.02$ (Gaussian-RFIM).

\begin{figure}[ht]
\begin{center}
\scalebox{0.5}[0.5]{\rotatebox{-90}
         {\includegraphics{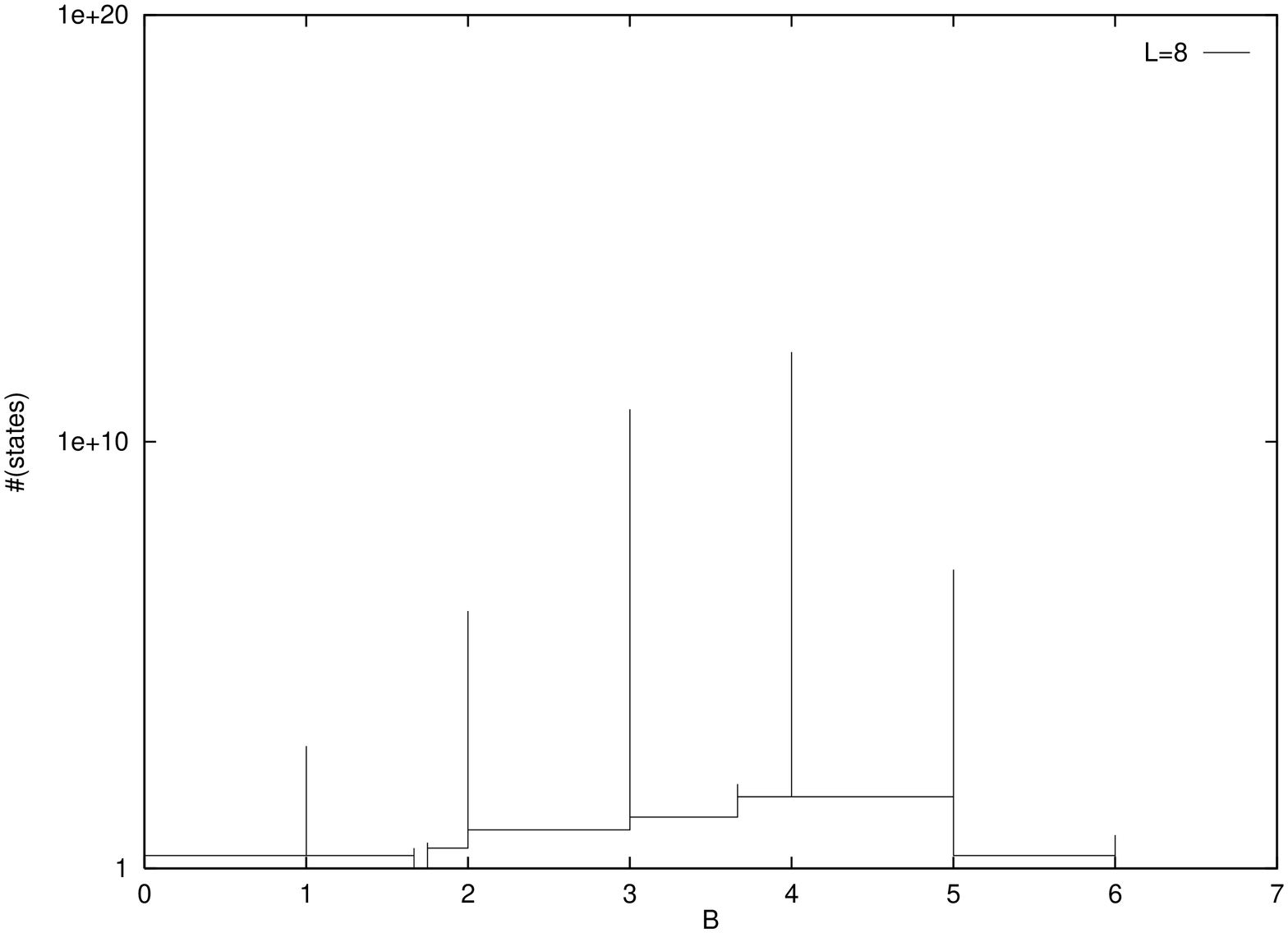}}}
\scalebox{0.5}[0.5]{\rotatebox{-90}
         {\includegraphics{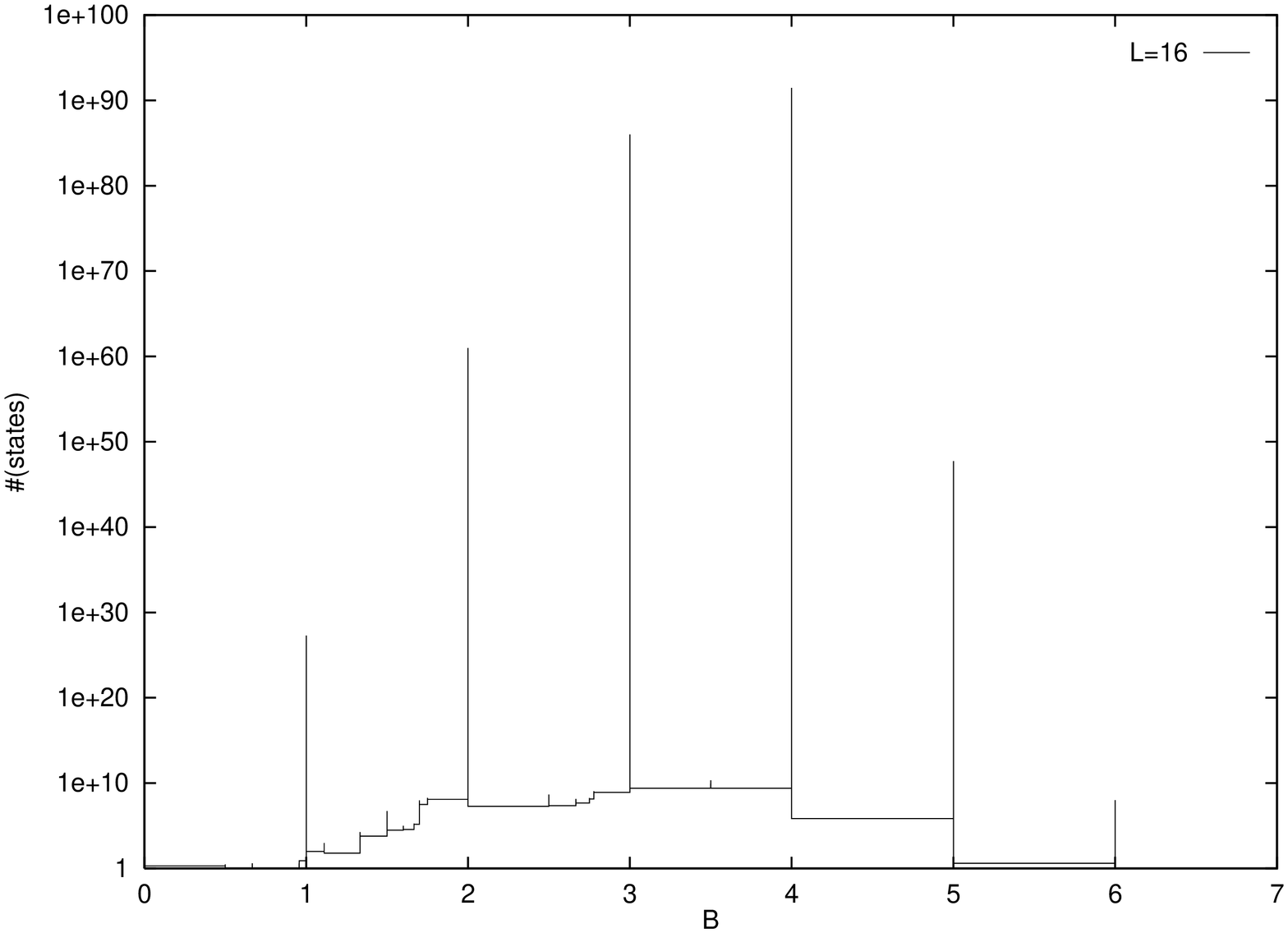}}}
\end{center}
\caption{}
\label{p_af_degdeg}
\end{figure}

In the last part we focus on the degree of degeneration (dod).
In figure \ref{p_af_degdeg} the dod
 of the sample DAFF systems is displayed as function
of the field $B$. The dod is not calculated exactly,
but estimated very accurately\footnote{For the graph types which occur here,
the estimate seems to be exact: For small systems the value is the same. For
large system the type of the graph is the same, as we will see.} 
by an algorithm found in \cite{degdeg}.

The degree of degeneracy is there very large, where the external field
takes integer values. For non integer values the dod takes the largest values
for $B \in (3,4)$. 

 We found that there are 
only small connected subgraphs in $\overline{R}$ of small height 
(the maximum length path in the graph) at fields $B$ where no jump occurs. 
So the spin-clusters
contributing to the degeneracy are flipping almost independently, 
i.e. the ground state structure is quite simple. That means that the number 
of states are not reduced very much from $2^{\#n}$, where $\#n$ is the
number of inner clusters.

We could find that the average 
dod can be approximated very good
by a function of $\#n$ and of the
number of edges $\#e$ in $\overline{R}$ over the full $B$-range:

\begin{equation}
\mbox{dod} = 2^{\#n}\alpha^{\#e}
\label{dod_formula}
\end{equation}

This means, that for $\#e =0$ all inner clusters are independent, so the
dod depends exponential on the number of nodes. The edges put constraints
on the nodes, so the number of possible states is reduced.
So it is possible to describe the ground state structure by a single value
$\alpha$. We measured $\alpha \approx 0.79 \pm 0.03$. This value is quite
independent of the system size and the external field\footnote{The distribution
of $\alpha$-values has a peak at $\alpha=0.75$. That is the value of a system
consisting only of independent clusters and 
of pairs of clusters each connected by 
one edge.}.

\begin{figure}[ht]
\begin{center}
\scalebox{0.5}[0.5]{\rotatebox{-90}
         {\includegraphics{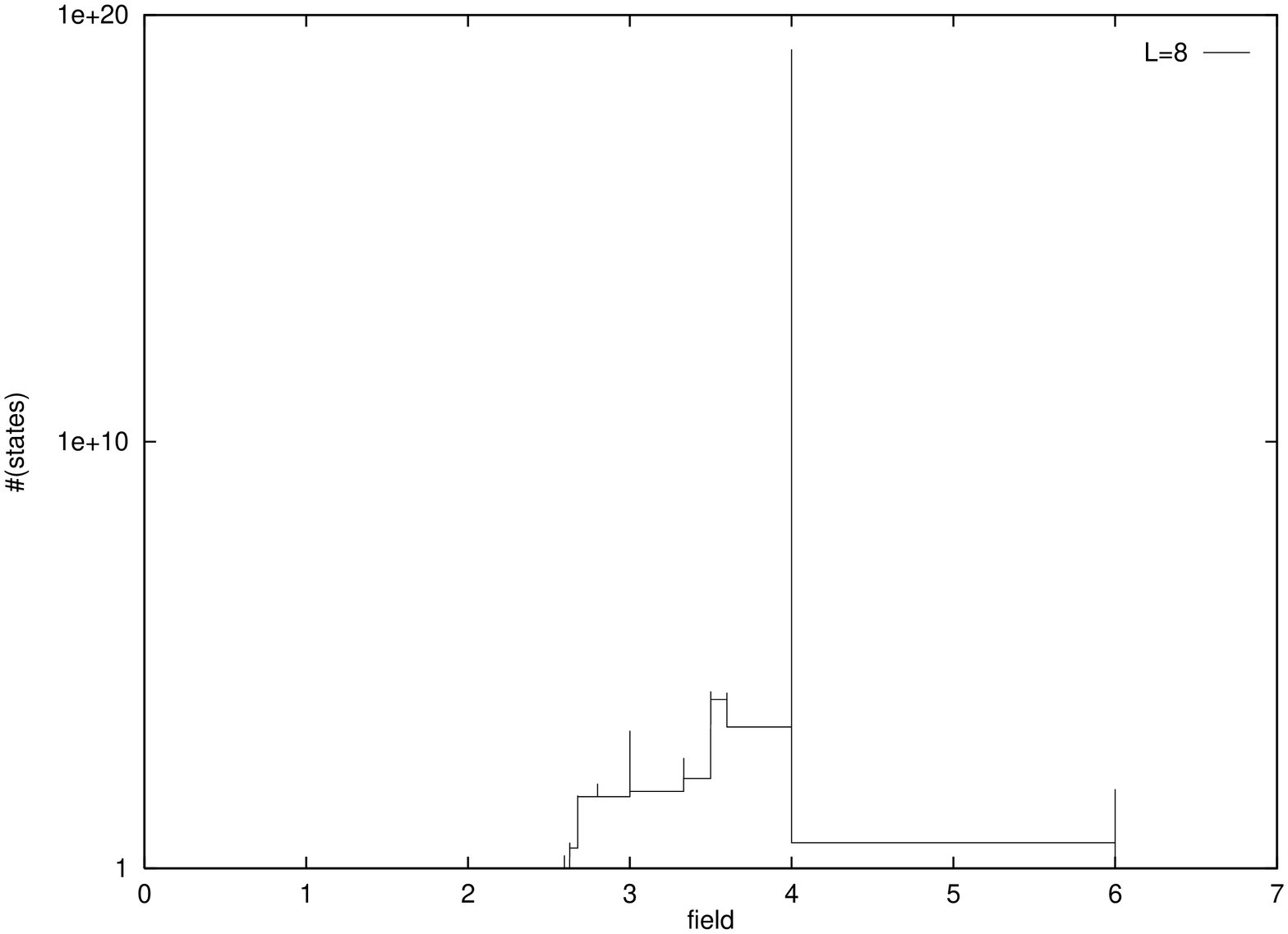}}}
\scalebox{0.5}[0.5]{\rotatebox{-90}
         {\includegraphics{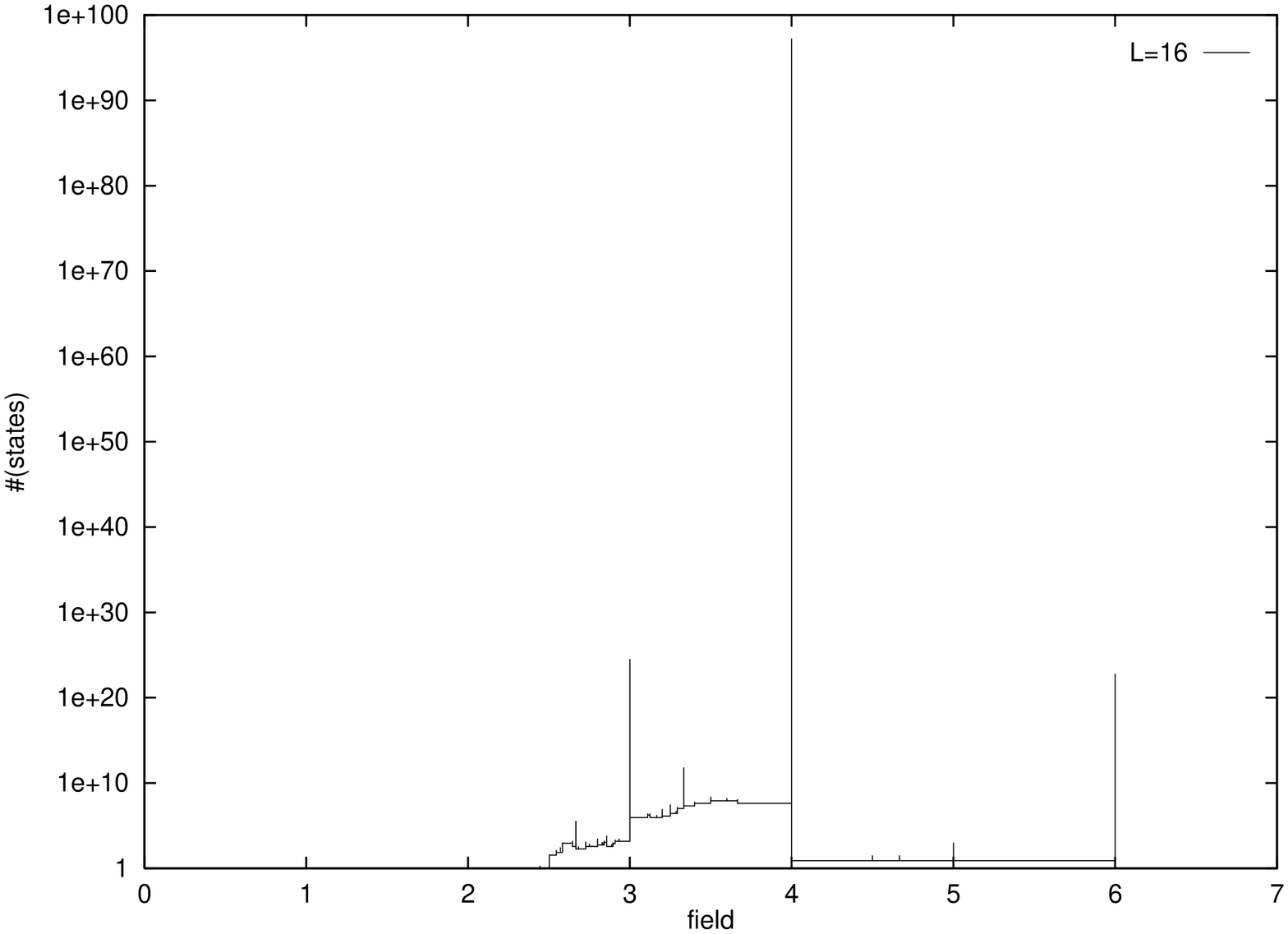}}}
\end{center}
\caption{}
\label{p_rf_degdeg}
\end{figure}

Figure \ref{p_rf_degdeg} shows the dod for the two sample $\pm \Delta$-RFIM
systems. Here the dod takes its largest values at the integer-fields
$\Delta = 3,4,6$ and in $(3,4)$. For the Gaussian-RFIM {\em no degeneration}
occurs except the twofold degeneration at the jumps. This is to be expected,
 because it is very
unlikely that the sum of the random fields $\Sigma_C$ is zero for a cluster
$C$. The dod of the $\pm\Delta$-RFIM can also be approximated very well by 
equation (\ref{dod_formula}). We got an average $\alpha=0.7582 \pm 0.0002$.

\begin{figure}[ht]
\begin{center}
\scalebox{0.5}[0.5]{\rotatebox{-90}{\includegraphics{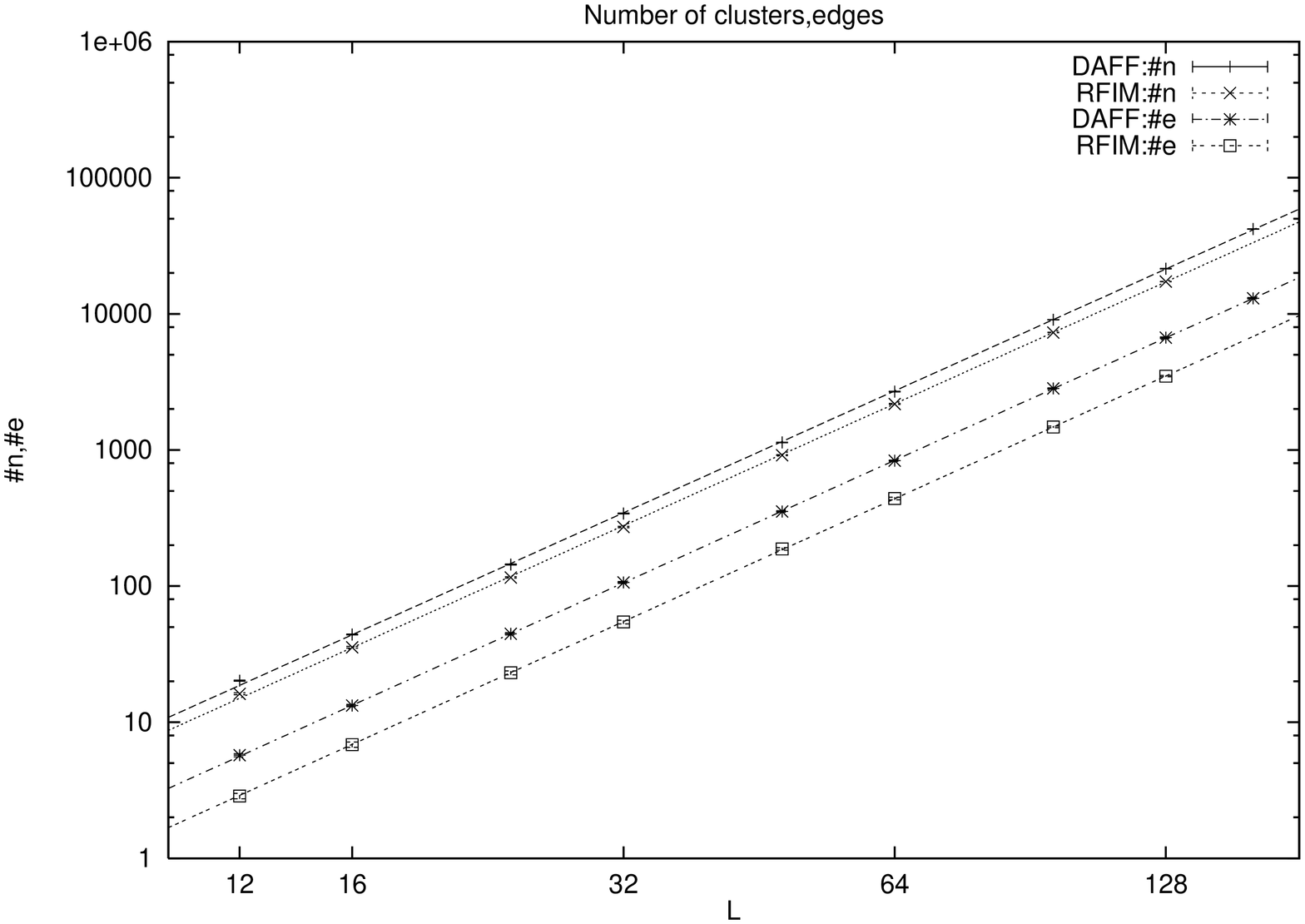}}}
\end{center}
\caption{}
\label{p_num_xxx}
\end{figure}

Since the dod is mainly determined by the number $\#n$ of (inner) clusters
and the number $\#e$ of edges, we calculated these values for the 
DAFF/$\pm\Delta$-RFIM at $B=\Delta=3.9999$ for sizes up to $L=160$ (from
2560 realizations for $L=12$ to 10 realizations for the largest systems).
The result is shown in figure \ref{p_num_xxx}.

For both system types all numbers have a $L^{3.00}$ dependence on $L$. The
number of edges is about 3 to four times smaller than the number of clusters.
For the DAFF about $4.6$ percent of the spins are contained 
in the inner clusters, for
the $\pm\Delta$-RFIM this value is only 1.8 percent. These values are
almost independent of the system size.
Because the number of clusters/state depend linearly on the system size, 
this shows again that the ground state of the DAFF/RFIM consist of many
independent small clusters. So the number of clusters increases, but not
the size of the clusters and not the degree of connection between the
clusters. This means that the degree of degeneracy depends
exponential on the size of the system.

\section{Conclusions}

In this paper we have presented an algorithm for calculating exact
ground states of random field systems and diluted antiferromagnets in
polynomial time for all fields $B,\Delta$. The complete degeneracy of
all states is representable by a relation of (anti-)ferromagnetic clusters.
The use of new fast algorithm allowed us to calculate systems of sizes
up to $160^3$.

We found that the transition to the domain states occurs for the DAFF
at $B_c<1$ in
contradiction to previous results. The result for the $\pm\Delta$-RFIM 
$\Delta_c=2.35\pm 0.01$ was consistent with previous calculations. For
the Gaussian RFIM we  found $\Delta_c=2.49\pm 0.04$.

The order parameters of the systems are stepwise constant functions of
the field. The number of jumps inside a given $B,\Delta$ interval
increases faster than $L$, indicating
that the infinite system has a rather smooth behavior.

The ground state degeneracy of the DAFF and the $\pm\Delta$-RFIM
is very high, especially when the field is
integer valued. Apart from these values the degree of degeneracy is
at highest in the domain state. 
The degree of degeneracy increases exponentially with
system size. The degeneration is rather simple, because the ground states
 consist
of many small, only loosely coupled clusters. Not more than 5 resp.\ 2 percent
of all spins contribute to the degeneracy. The Gaussian-RFIM is
not degenerated at all, except at the jump fields, where one cluster
can take two orientations.
 
Using the methods presented and referenced in this paper many more
information about the ground state structure of DAFFs and RFIMs will
be obtained in the future. A detailed analysis of the numbers, sizes and
forms of the clusters forming the ground states
and type of connections between them is possible. 
Because the systems get smoother with increasing size,
finite size scaling techniques should be applicable, to gain more insight into
the transition from the long range order to the domain state.

\section{Acknowledgments}
\begin{sloppypar}
We would like to thank D.W. Heermann for manifold support and
K.D. Usadel and J. Esser for fruitful discussions and critical reading 
of the manuscript.
We thank the {\em Paderborn Center for Parallel Computing}, the 
{\em Uni\-ver\-si\-t\"ats\-rechen\-zen\-trum Hei\-del\-berg} and the 
{\em Rechenzentrum Uni\-ver\-si\-t\"at
Duisburg} for the allocation of computer time. This work was supported
by the Graduiertenkolleg ``Modellierung und Wissenschaftliches Rechnen in 
Mathematik und Naturwissenschaften'' at the
{\em In\-ter\-diszi\-pli\-n\"a\-res Zentrum f\"ur Wissenschaftliches Rechnen}
 in Heidelberg. 
\end{sloppypar}

{\Large \bf Captions}
\begin{enumerate}
\item Example RFIM system with four spins.

\item Ground state of example RFIM given by the relation $\quer{R}$.

\item Ground state $\quer{R}$ of sample DAFF ($L=10$, $p=0.5$) 
for $9/7<B<1.5$.

\item Ground state $\quer{R}$ of sample DAFF ($L=10$, $p=0.5$) 
for $B=1.5$.

\item Ground state $\quer{R}$ of sample DAFF ($L=10$, $p=0.5$) 
for $1.5<B<1.6$.

\item Antiferromagnetic order parameter of sample DAFFs ($L=8$, $p=0.5$)
and ($L=16$, $p=0.5$.) 
for $0 \leq B <  6.5$.

\item Barrier field $B_{a<a_0}(1/L)$ for $L=8,\ldots,48$ 
($a_0 =0.2,0.5,0.8$) and fit functions $B_{a<a_0}(1/L)=c/L+B_{a<a_0}(0)$.
\item Antiferromagnetic order parameter of sample RFIMs ($L=8$)
and ($L=16$) 
for $0 \leq \Delta <  6.5$.

\item Barrier field $\Delta_{m<m_0}(1/L)$ for $L=8,\ldots,32$ 
($m_0 =0.2,0.5,0.8$) and fit functions 
$\Delta_{m<m_0}(1/L)=c/L+\Delta_{m<m_0}(0)$.

\item Average number of jumps in $B \in [1,2.8]$ (Gaussian-RFIM) resp. 
$[0,6]$ ($\pm \Delta$-RFIM, DAFF) 
as function of system size in a double logarithmic scale and fit functions 
$\#j(L)=cL^b$.

\item Degree of degeneracy of sample DAFFs ($L=8$, $p=0.5$)
and ($L=16$, $p=0.5$) 
for $0 \leq B <  6.5$.

\item Degree of degeneracy of sample RFIMs ($L=8$)
and ($L=16$) 
for $0 \leq \Delta <  6.5$.

\item Number of clusters $\#n$, edges $\#e$ as function of system size 
$L$ for
the DAFF and the $\pm\Delta$-RFIM together with fits of form $f(L)=cL^b$.

\end{enumerate}

\begin{figure}[ht]
\begin{center}
\begin{picture}(6,1)
\multiput(0,0)(0,1){2}{\line(1,0){6}}
\multiput(0,0)(1.5,0){5}{\line(0,1){1}}
\multiput(0.225,0.25)(0,0.5){2}{\line(1,0){0.3}}
\multiput(0.375,0.1)(0,0.5){2}{\line(0,1){0.3}}	                    
\multiput(4.725,0.25)(0,0.5){2}{\line(1,0){0.3}}
\put(1.125,0.2){\vector(0,1){0.6}}
\put(5.625,0.8){\vector(0,-1){0.6}}
\put(2.25,0.0){\makebox(0.75,1){\Large$\sigma_2$}}
\put(3.75,0.0){\makebox(0.75,1){\Large$\sigma_3$}}
\end{picture}
\end{center}
\caption{}
\label{example_sys}
\end{figure}

\end{document}